\RequirePackage{fix-cm}

\documentclass[smallextended]{svjour3}       % onecolumn (second format)
\usepackage[left=3.17cm,right=3.17cm,top=2.54cm,
headheight=0.5cm,headsep=0.54cm,bottom=2.54cm,footskip=0.79cm
]{geometry}

\usepackage{amsmath,amssymb}

\usepackage{graphicx}
\begin{document}

\title{Lower bound of multipartite concurrence based on sub-partite quantum systems}%\thanks{Grants or other notes



\author{Wei Chen         \and
        Xue-Na Zhu        \and
        Shao-Ming Fei     \and Zhu-Jun Zheng%etc.
}


\institute{Wei Chen  \at
              School of Computer Science and Network Security, Dongguan University of Technology, Dongguan 523808, P.R.China \\
              \email{auwchen@scut.edu.cn}           %  \\
           \and
           Xue-Na Zhu \at
               Department of Mathematics and Statistics Science, Ludong University, Yantai 264025, P.R.China
           \and
           Shao-Ming Fei  \at
              School of Mathematical Sciences, Capital Normal University, Beijing 100048, China
              \and
           Shao-Ming Fei  \at
              Max-Planck-Institute for Mathematics in the Sciences, Leipzig 04103, Germany
              \and
           Zhu-Jun Zheng  \at
           School of Mathematics, South China University of Technology,
Guangzhou 510641, China
}

\date{Received: date / Accepted: date}

\maketitle

\begin{abstract}
We study the concurrence of arbitrary dimensional multipartite quantum systems. An explicit analytical lower bound of concurrence for four-partite mixed states is obtained in terms of the concurrences of tripartite mixed states. Detailed examples are given to show that our lower bounds improve the existing lower bounds of concurrence. The approach is generalized to five-partite quantum systems.
 %\PACS{03.67.Mn \and 03.65.Ud }
\end{abstract}

\section{Introduction}\label{sec1}
\label{intro}
As a striking feature of quantum physics and an essential resource in quantum information processing \cite{Nie}-\cite{Wer},
quantum entanglement has attracted much attention in recent years \cite{Hor}-\cite{Vic}.
Its potential applications in quantum information processing have been demonstrated in, such as quantum computation \cite{di},
quantum teleportation \cite{teleportation}, dense coding
\cite{dense}, quantum cryptographic schemes \cite{schemes},
entanglement swapping \cite{swapping}, remote states preparation \cite{RSP1}, and in many pioneering experiments.

To give a proper description and qualify the quantum entanglement for a given quantum state, many entanglement measures
have been introduced, such as the entanglement of formation \cite{Ben2} for bipartite quantum systems and concurrence \cite{Uhl} for any multipartite quantum systems.
For the two qubit case, the entanglement of formation is proven to be a monotonically increasing function of the concurrence and an
elegant formula for the concurrence was derived analytically by Wootters \cite{Woo}. However, except for bipartite qubit systems and some special symmetric states \cite{Ter},
there have been no explicit analytic formulas of concurrence for arbitrary high-dimensional mixed states,
due to the extremizations involved in the computation.
Instead of analytic formulas, some progress has been made toward the analytical lower bounds of concurrence.
A lower bound of concurrence based on local uncertainty relation criterion is derived in \cite{Vic}.
This bound is further optimized in \cite{Zha}. For arbitrary bipartite quantum states, Refs \cite{Ger}-\cite{Fan} provide a
detailed proof of an analytical lower bound of concurrence in terms of a different approach that has a close relationship with
the distillability of bipartite quantum states.

In \cite{Fan}-\cite{Zhu}, the authors presented a lower bound of concurrence by
decomposing the joint Hilbert space into many $2\otimes 2$ and $s\otimes t$-dimensional subspaces, which improve all the known lower bounds of concurrence. A similar nice algorithms and progress have been made towards
lower bounds of concurrence for tripartite quantum systems \cite{Zhu0,Ch} and other multipartite quantum
systems \cite{Wang}-\cite{Zhu1} by bipartite partitions of the whole quantum system. One would like to ask naturally if it is possible to improve further the lower bound of concurrence by using tripartite and $M$-partite concurrences of an $N$-partite ($M < N$) systems.

In this paper, we first provide lower bounds of concurrence for arbitrary dimensional four-partite systems in terms of tripartite concurrences.
Detailed examples are given to show that these bounds are better than the well known existing lower bounds of concurrence.
We then generalize lower bound of concurrence to arbitrary multipartite case.
\medskip
\section{Lower bounds of concurrence for four-partite mixed states}\label{sec2}
We first recall the definition and some lower bounds of the multipartite concurrence.
Let $H_i$, $i=1,\cdots,N$, be $d_i$ dimensional Hilbert spaces.
The concurrence of an $N$-partite pure state $|\psi\rangle\in H_1\otimes H_2\otimes\cdots \otimes H_N$ is defined by \cite{Aol},
\begin{equation}\label{1}
C_N(|\psi\rangle) = 2^{1-\frac{N}{2}}\sqrt{(2^N-2)-\sum_{\alpha} Tr[\rho_{\alpha}^{2}]},
\end{equation}
where the index $\alpha$ labels all $2^N-2$ non-trivial subsystems of the $N$-partite quantum systems and
$\rho_{\alpha}$ are the corresponding reduced density matrices.

% If we list all the $2^N-2$ reduced density matrices in the following way:
%$\{\rho_1, \rho_2, \cdots, \rho_N, \rho_{12}, \rho_{13}, \cdots, \rho_{1N}, \rho_{23}, \cdots, \rho_{N-1N},$
%$\cdots, \rho_{12\cdots N-1}, \cdots, \rho_{23\cdots N}\}.$ (\ref{1}) can be reexpressed as

%\begin{equation}
%C_N(|\psi\rangle) = 2^{1-\frac{N}{2}}\sqrt{(2^N-2)-\sum_{k=1}^{2^N-2} Tr[\rho_{k}^{2}]},
%\end{equation}

For a mixed multipartite quantum state $\rho = \sum_{i}p_{i}|\psi_{i}\rangle\langle\psi_{i}| \in H_1\otimes H_2\otimes\cdot\cdot\cdot \otimes H_N$,
$p_{i}\geq 0$, $\sum_{i}p_{i} = 1$, the concurrence is given by the convex roof:
\begin{equation}\label{2}
C_N(\rho) = \min_{\{p_{i},|\psi_{i}>\}}\sum_{i}p_{i}C_N(|\psi_{i}\rangle),
\end{equation}
where the minimum is taken over all possible convex partitions of $\rho$ into pure state ensembles $\{|\psi_{i}\rangle\}$
with probability distributions $\{p_{i}\}$.

In \cite{Wang} the authors obtained lower bounds of multipartite concurrence in terms of the concurrences of
bipartite partitioned states of the whole quantum system. For an $N$-partite quantum
pure state $|\psi\rangle\in H_1\otimes H_2\otimes\cdots \otimes H_N$, $dim H_i = d_{i}$, $i=1,\cdots, N$,
the concurrence of bipartite partition between the subsystems $1 2 \cdots M$ and $M+1\cdots N$ is defined by
\begin{eqnarray}\label{3}
C_2(|\psi\rangle\langle\psi|) = \sqrt{2(1-Tr[\rho^{2}_{1 2\cdots M}])},
\end{eqnarray}
where $\rho_{1 2\cdots M} = Tr_{M+1\cdots N}\{|\psi\rangle\langle\psi|\}$
is the reduced density matrix of $\rho = |\psi\rangle\langle\psi|$ by tracing over the subsystems $M+1\cdots N.$
For a mixed multipartite quantum state $\rho = \sum_{i}p_i |\psi_{i}\rangle\langle\psi_{i}| \in  H_1\otimes H_2\otimes\cdots \otimes H_N,$
the corresponding concurrence $C_2(\rho)$ is given by the convex roof:
\begin{eqnarray}\label{4}
C_2(\rho) = \min_{\{p_{i},|\psi_{i}\rangle\}}\sum_{i}p_{i}C_2(|\psi_{i}\rangle\langle\psi_{i}|).
\end{eqnarray}
A relation between the concurrence (\ref{2}) and the bipartite concurrence (\ref{4}) has been presented in \cite{Wang}:
For a multipartite quantum state $\rho \in  H_1\otimes H_2\otimes\cdots \otimes H_N$ with $N\geq 3,$ the following inequality holds,
\begin{eqnarray}\label{5}
C_N(\rho) \geq \max 2^{\frac{3-N}{2}}C_2(\rho),
\end{eqnarray}
where the maximum is taken over all kinds of bipartite concurrences.

In terms of the lower bounds of bipartite concurrence, in \cite{Zhu1}
further relations between the concurrence (\ref{2}) and the bipartite concurrence (\ref{4}) has been obtained:

\begin{eqnarray}\label{6}
C_N(\rho) \geq \max_{M = 1, 2, \cdots, N-1} \{2^{\frac{1-N}{2}}\sqrt{2^{N-M}+2^M-2}C_2(\rho_M)\}
\end{eqnarray}

for $N\geq 3,$ where the maximum is taken over all kinds of bipartite concurrences for given $M$.
In particularly, if $N=3,$ one has $C_3(\rho) \geq \max \{C_2(\rho_1), C_2(\rho_2)\}$. If $N=4,$ one
 gets $C_4(\rho) \geq \max \{C_2(\rho_1), \frac{\sqrt{3}}{2}C_2(\rho_2), C_2(\rho_3)\}.$

For multi-qubit systems, in \cite{Zhu2} the authors get the analytical lower bounds in terms of the monogamy inequality:
For any four-qubit mixed quantum state $\rho,$ the concurrence $C(\rho)$ satisfies
\begin{eqnarray}\label{7}
C_4^{2}(\rho) \geq \sum^{3}_{i=1}\sum^{4}_{j>i}(T_i+T_j)C_{ij}^{2}(\rho),
\end{eqnarray}
where

$$T_1 = 1+\{-\frac{2-x}{2}|\frac{2-x}{2}\}+\{-\frac{2-y}{2}|\frac{2-y}{2}\}+\{-\frac{2-z}{2}|\frac{2-z}{2}\},$$
$$T_2 = 1+\{\frac{2-x}{2}|-\frac{2-x}{2}\}+\{-\frac{y}{2}|\frac{y}{2}\}+\{-\frac{z}{2}|\frac{z}{2}\},$$
$$T_3 = 1+\{-\frac{x}{2}|\frac{x}{2}\}+\{\frac{2-y}{2}|-\frac{2-y}{2}\}+\{\frac{z}{2}|-\frac{2-z}{2}\},$$
$$T_4 = 1+\{\frac{x}{2}|-\frac{x}{2}\}+\{\frac{y}{2}|-\frac{y}{2}\}+\{\frac{2-z}{2}|-\frac{2-z}{2}\},$$

\noindent$x, y, z \in [0,2],$ the bracket $\{a|b\}$ is defined such that one may either take the first element $a$ or the second element
$b$ from $\{a|b\}$, and for example $C_{12}^{2}(\rho)$ denotes the concurrence of the reduced state $\rho_{12}=Tr_{34}(\rho)$. However, for any given pair $a$ and $b$, once the first (the second) has been taken, then
in a formula one always takes the first (the second) element in all the following brackets containing the same two elements $a$ and $b$.

In order to improve the lower bounds of concurrence, in the following we consider tripartite concurrence $C_{3}(\rho)$, instead of the bipartite concurrence $C_{2}(\rho)$.
For an $N$-partite quantum pure state $|\psi\rangle \in H_1\otimes H_2\otimes\cdots \otimes H_N$, $dim H_i = d_i$, $i = 1,2,\cdots N$ $(N\geq 3)$, we denote $M$ decomposition among subsystems $\{i^1\}, \{i^2\}, \cdots,\{i^{M_1}\}, \{k_1^1, k_2^1\}, \{k_1^2, k_2^2\}, \cdots, \{k_1^{M_2}, k_2^{M_2}\},\cdots,\{q_1^{1},\cdots, q_j^{1}\},\{q_1^{2},$
\noindent$\cdots, q_j^{2}\}, \cdots, \{q_1^{M_j},\cdots, q_j^{M_j}\},$ where
$\{i^1, i^2, \cdots, i^{M_1}, ~k_1^1, ~k_2^1, ~k_1^2, ~k_2^2, \cdots, k_1^{M_2}, k_2^{M_2},\cdots,$ $q_1^{1},\cdots, q_j^{1},$
\noindent$\cdots, q_1^{M_j},\cdots, q_j^{M_j}\} = \{1,2,\cdots,N\}$ and $\sum_{k=1}^{j}M_{k}=M,$ $\sum_{k=1}^{j}kM_{k}=N,$ the concurrence of $M-$partite decomposition among the above subsysytems is given by

\begin{eqnarray}\label{8}
C_{M}(|\psi\rangle\langle\psi|) = 2^{1-\frac{M}{2}}\sqrt{(2^M-2)-\sum_{\alpha}Tr[\rho_{\alpha}^{2}]},
\end{eqnarray}
where $\emptyset \neq \alpha \subsetneq \{\{i^1\}, \{i^2\}, \cdots,\{i^{M_1}\}, \{k_1^1, k_2^1\}, \{k_1^2, k_2^2\},$ $ \cdots,\{k_1^{M_2}, k_2^{M_2}\},\cdots,\{q_1^{1},\cdots, q_j^{1}\},\cdots,$
\noindent$ \{q_1^{M_j},\cdots, q_j^{M_j}\}\}$ and $\rho_{\alpha}$ are the corresponding reduced density matrices.

For example, we can define the concurrence of tripartite decomposition among subsystems $1,2,\cdots, M$, $M+1, \cdots, L$ and $L+1, \cdots, N$ as,

\begin{eqnarray}\label{9}
C_3(|\psi\rangle\langle\psi|) = \sqrt{3-Tr[\rho_{12\cdots M}^2+\rho_{M+1\cdots L}^2+\rho_{L+1\cdots N}^2]},
\end{eqnarray}

\noindent where $\rho_{12\cdots M} = Tr_{M+1, \cdots, L, L+1, \cdots, N}(|\psi\rangle\langle\psi|)$ is the reduced density matrix of
$\rho = |\psi\rangle\langle\psi|$ by tracing over the subsystems $M+1, \cdots, L, L+1, \cdots, N$. Similar definitions apply
to $\rho_{M+1\cdots L}$ and $\rho_{L+1\cdots N}$. The rearrangement of the subsystems are implied naturally, so if take $N=4, M=3,$ there are six different  partitions of four system: $1|2|34, 1|3|24, 1|4|23, 12|3|4, 13|2|4, 14|2|3,$ then we can get the following theorem:

\noindent{\bf Theorem 1.} For a multipartite quantum state $\rho \in  H_1\otimes H_2\otimes H_3 \otimes H_4,$  then the following inequality holds,
\begin{eqnarray}\label{10}
C_4^{2}(\rho) \geq  \widetilde{C_3}^{2} (\rho),
\end{eqnarray}
where $\widetilde{C_3}^{2} (\rho) = \frac{1}{6}(C_3^{2}(\rho_{1|2|34})+C_3^{2}(\rho_{1|3|24})+C_3^{2}(\rho_{1|4|23})+C_3^{2}(\rho_{12|3|4})+C_3^{2}(\rho_{13|2|4})+C_3^{2}(\rho_{14|2|3})).$

\noindent \emph{Proof:} For a pure multipartite state $|\psi\rangle \in H_1\otimes H_2\otimes H_3 \otimes H_4,$ let $\rho = |\psi\rangle\langle\psi|,$
From (1), we have
\begin{eqnarray}\label{11}
C_4^{2}(\rho)=\frac{1}{2} (\sum_{i=1}^{4}(1-tr\rho^{2}_{i}) + \sum_{i=2}^{4}(1-tr\rho^{2}_{1i}))
\end{eqnarray}
and
\begin{eqnarray}\label{12}
C_3^{2}(\rho_{i|j|kl})= (1-tr\rho^{2}_{i}) + (1-tr\rho^{2}_{j}) + (1-tr\rho^{2}_{kl}),
\end{eqnarray}
where $\rho_{i}=Tr_{jkl}(\rho), \rho_{j}=Tr_{ikl}(\rho), \rho_{kl}=Tr_{ij}(\rho).$

Then from (\ref{11}) and (\ref{12}), we have $C_4^{2}(\rho) \geq  \frac{1}{6}(C_3^{2}(\rho_{1|2|34})+C_3^{2}(\rho_{1|3|24})+C_3^{2}(\rho_{1|4|23})+C_3^{2}(\rho_{12|3|4})+C_3^{2}(\rho_{13|2|4})+C_3^{2}(\rho_{14|2|3})).$

Assuming that a mixed state $\rho = \sum_{i}p_{i}|\psi_{i}\rangle\langle\psi_{i}|$ attains the minimal partition of the multipartite concurrence, one has,

$$C_{4}^{2}(\rho) = (\sum_{i}p_{i}C_{4}(|\psi_{i}\rangle\langle\psi_{i}|))^{2}~~~~~~~~~~~~~~~~~~~~~~~~~~~~~~~~~~~~~~~~~~~~~~~~~~~~~~~~~~~~~~~~~~~~~~~~~$$
$$~~~~~~\geq (\sum_{i}p_{i}\sqrt{\frac{1}{6}(C_3^{2}((|\psi_{i}\rangle)_{1|2|34})+C_3^{2}((|\psi_{i}\rangle)_{1|3|24})+\cdots+C_3^{2}((|\psi_{i}\rangle)_{14|2|3}))}~)^{2}~~~~~~~~~~~~~~~~~~~~~~~~$$
$$~~~~~~~~~~\geq (\sum_{i}p_{i}\frac{1}{\sqrt{6}}C_3((|\psi_{i}\rangle)_{1|2|34}))^{2} + (\sum_{i}p_{i}\frac{1}{\sqrt{6}}C_3((|\psi_{i}\rangle)_{1|3|24}))^{2} + \cdots + (\sum_{i}p_{i}\frac{1}{\sqrt{6}}C_3((|\psi_{i}\rangle)_{14|2|3}))^{2}$$
$$~\geq \frac{1}{6}(C_3^{2}(\rho_{1|2|34})+C_3^{2}(\rho_{1|3|24})+C_3^{2}(\rho_{1|4|23})+C_3^{2}(\rho_{12|3|4})+C_3^{2}(\rho_{13|2|4})+C_3^{2}(\rho_{14|2|3})),~~~$$

\noindent where the relation $(\sum_{j}(\sum_{i}x_{ij})^{2})^{\frac{1}{2}}\leq \sum_{i}(\sum_{j}x_{ij}^{2})^{\frac{1}{2}}$ has been used in second inequality. Therefore, we have (\ref{10}).

\qed
In order to show that our lower bound (\ref{10}) can detect entanglement better, let us consider the following examples.

\noindent{\bf Example 1.} We now first consider a simple case, the generalized four-qubit GHZ state: $|\psi\rangle = cos\theta|0000\rangle + sin\theta|1111\rangle$. We have $C_4(|\psi\rangle) = \sqrt{7sin^2\theta cos^2\theta}.$
From our lower bound (\ref{10}), we have
$C_{4}(\rho) \geq \sqrt{6sin^2\theta cos^2\theta}$,
which is generally greater than the bounds $\sqrt{4sin^2\theta cos^2\theta}$ from \cite{Zhu1} and $\sqrt{2sin^2\theta cos^2\theta}$ from \cite{Wang}.

\noindent{\bf Example 2.} Now consider the quantum mixed state
$\rho = \frac{1-t}{16}I_{16} + t|\phi\rangle\langle\phi|,$ with $|\phi\rangle = \frac{1}{2}(|0000\rangle + |0011\rangle + |1100\rangle + |1111\rangle)$,
where $I_{16}$ denotes the $16\times 16$ identity matrix. By Theorem 4 in \cite{Ch}, we obtain
\[C^2(\rho_{12|3|4})\geq \left\{\begin{array}{ll}
~~~~~~0,&\text{$0\leq t \leq\frac{1}{9}$ },\\
\frac{81t^{2}-18t+1}{192},&\text{$\frac{1}{9}< t \leq\frac{1}{5}$},\\
\frac{181t^{2}-58t+5}{192},&
\text{$\frac{1}{5}< t \leq 1$}.
\end{array}\right.\]

Also we can get \[C^2(\rho_{1|3|24})\geq \left\{\begin{array}{ll}
~~~~~~0,&\text{$0\leq t \leq\frac{1}{5}$ },\\
\frac{175t^{2}-70t+7}{192},&
\text{$\frac{1}{5}< t \leq 1$}.
\end{array}\right.\]

Similarly, $C^2(\rho_{1|2|34})$ has the same lower bound as $C^2(\rho_{12|3|4})$, and $C^2(\rho_{1|4|23}), C^2(\rho_{13|2|4}),$
$C^2(\rho_{14|2|3})$ have the same lower bound as $C^2(\rho_{1|3|24}).$ Associated with (10), we have

\[C^2_{4}(\rho)\geq \left\{\begin{array}{ll}
~~~~~~0,&\text{$0\leq t \leq\frac{1}{9}$ },\\
\frac{81t^{2}-18t+1}{576},&\text{$\frac{1}{9}< t \leq\frac{1}{5}$},\\
\frac{531t^{2}-198t+19}{576},&
\text{$\frac{1}{5}< t \leq 1$}.
\end{array}\right.\]

So our result can detect the entanglement of $\rho$ when
$\frac{1}{9}< t \leq 1,$ see FIG.1. While the lower bound of Theorem 1 in \cite{Zhu2} is $C^2(\rho)\geq 0,$ when $\frac{1}{9}< t \leq \frac{1}{3}$, which can not detect the entanglement of the above $\rho.$ Also we can found that our lower bound are larger than the lower bound of Theorem 1 in \cite{Zhu2} when $\frac{1}{9}< t \leq \frac{111+4\sqrt{106}}{255},$

\begin{figure}[htpb]\label{Fig1}
\centering\
\includegraphics[width=7cm, height=5cm]{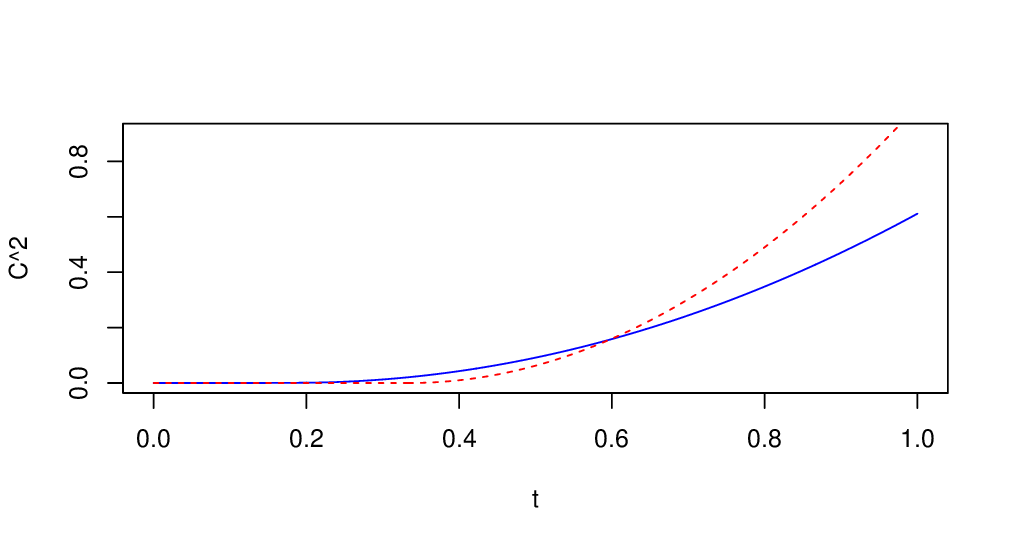}
\caption{Solid line  for the lower bound from (\ref{10}), which detects the entanglement of $\rho$ when $\frac{1}{9}< t \leq 1$.
 Dashed line for the lower bound from Theorem 1 in \cite{Zhu2}. It detects entanglement only for $\frac{1}{3} <t \leq 1 $.}
\end{figure}

Similarly, the lower bound of Theorem 1 in \cite{Qi} is $C^2(\rho)\geq 0,$ when $\frac{1}{9}< t \leq \frac{1}{3}$, which can not detect the entanglement of the above $\rho.$ Also we can found that our lower bound are larger than the lower bound of Theorem 1 in \cite{Qi} when $\frac{1}{9}< t \leq \frac{219+4\sqrt{187}}{579},$ see FIG.2.

\begin{figure}[htpb]\label{Fig2}
\centering\
\includegraphics[width=7cm, height=5cm]{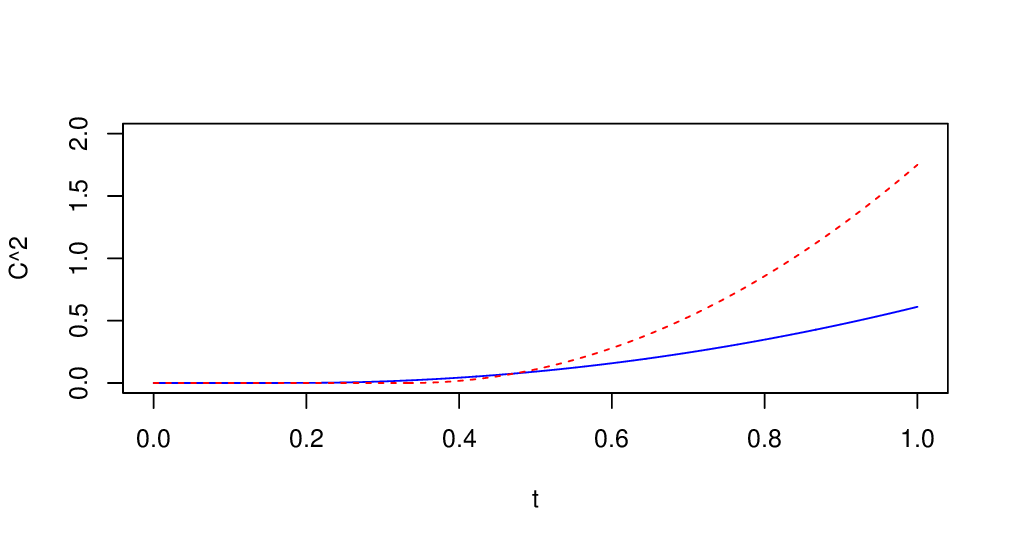}
\caption{Solid line  for the lower bound from (\ref{10}), which detects the entanglement of $\rho$ when $\frac{1}{9}< t \leq 1$.
 Dashed line for the lower bound from Theorem 1 in \cite{Qi}. It detects entanglement only for $\frac{1}{3} <t \leq 1$.}
\end{figure}

\noindent{\bf Example 3.} Let us consider the four-qubit Dicke state with two excitations\cite{Kie},
$$\rho = \frac{1-t}{16}I_{16} + t|D_4^{(2)}\rangle\langle D_4^{(2)}|,$$
where $|D_4^{(2)}\rangle = (|0011\rangle + |0101\rangle + |0110\rangle + |1001\rangle + |1010\rangle + |1100\rangle)/\sqrt{6}.$ When $\frac{3(8\sqrt{2}-3)}{119}< t \leq \frac{1}{3}$, we find that
the lower bound in \cite{Zhu2} and \cite{Qi} is $C^2(\rho)\geq 0,$  which can not detect entanglement. While our lower bound can detect entanglement when $\frac{3(8\sqrt{2}-3)}{119} < t \leq 1.$ So our bound can detect entanglement better.

\noindent{\bf Remark 1.} The definition of concurrence in \cite{Zhu2} is different from (1) up to a constant factor $2^{1-N/2}.$
In above examples and \cite{Qi}, the difference of the constant factor in defining the concurrence for pure states has already been taken
into account.

\section{Results of lower bounds of concurrence for arbitrary multipartite quantum systems}\label{sec3}

If we take $N=5, M=3,$ there are twenty-five different  partitions of five system:
$1|2|345, 1|3|245,$
\noindent$ 1|4|235, 1|5|234, 1|23|45, 1|24|35,1|25|34, 12|3|45, 12|34|5, 12|4|35, 13|2|45, 13|24|5,
13|4|25, 14|2|35,$
\noindent $14|23|5, 14|3|25, 15|2|34, 15|23|4, 15|3|24, 123|4|5, 134|2|5, 124|3|5, 135|2|4, 125|3|4, 145|2|3,$ then we have

\noindent{\bf Theorem 2.} For a multipartite quantum state $\rho \in  H_1\otimes H_2\otimes H_3 \otimes H_4\otimes H_5,$  then the following inequality holds,
\begin{eqnarray}\label{13}
C_5^{2}(\rho) \geq  \widetilde{C_3}^{2} (\rho),
\end{eqnarray}
where $\widetilde{C_3}^{2} (\rho) = \frac{1}{25}(C_3^{2}(\rho_{1|2|345})+C_3^{2}(\rho_{1|3|245})+\cdots + C_3^{2}(\rho_{145|2|3})).$

\noindent \emph{Proof:}  For a pure multipartite state $|\psi\rangle \in H_1\otimes H_2\otimes H_3 \otimes H_4\otimes H_5,$ let $\rho = |\psi\rangle\langle\psi|,$
From (\ref{1}), we have

\begin{eqnarray}\label{14}
C_5^{2}(\rho)=\frac{1}{4} (\sum_{i=1}^{5}(1-tr\rho^{2}_{i}) + \sum_{i=2}^{5}(1-tr\rho^{2}_{1i})+ \sum_{i=3}^{5}(1-tr\rho^{2}_{2i})+ \sum_{i=4}^{5}(1-tr\rho^{2}_{3i})+ (1-tr\rho^{2}_{45})),
\end{eqnarray}
\begin{eqnarray}\label{15}
C_3^{2}(\rho_{i|jt|kl})= (1-tr\rho^{2}_{i}) + (1-tr\rho^{2}_{jt}) + (1-tr\rho^{2}_{kl}),
\end{eqnarray}
where $\rho_{i}=Tr_{jtkl}(\rho), \rho_{jt}=Tr_{ikl}(\rho), \rho_{kl}=Tr_{ijt}(\rho),$
and
\begin{eqnarray}\label{16}
C_3^{2}(\rho_{i|j|kls})= (1-tr\rho^{2}_{i}) + (1-tr\rho^{2}_{j}) + (1-tr\rho^{2}_{kls}),
\end{eqnarray}
where $\rho_{i}=Tr_{jkls}(\rho), \rho_{j}=Tr_{ikls}(\rho), \rho_{kls}=Tr_{ij}(\rho).$

For a bipartite density matrix $\rho \in H_A\otimes H_B,$ from \cite{Zha}, one has
\begin{eqnarray}\label{17}
1-Tr(\rho^{2}_{AB})\leq (1-Tr(\rho^{2}_{A}))+(1-Tr(\rho^{2}_{B})),
\end{eqnarray}
where $\rho_{A}=Tr_{B}(\rho_{AB}), \rho_{B}=Tr_{A}(\rho_{AB}).$

Then from (\ref{14}), (\ref{15}), (\ref{16}) and (\ref{17}), we have $C_5^{2}(\rho) \geq \frac{1}{25}(C_3^{2}(\rho_{1|2|345})+C_3^{2}(\rho_{1|3|245})+\cdots + C_3^{2}(\rho_{145|2|3})).$

Assuming that a mixed state $\rho = \sum_{i}p_{i}|\psi_{i}\rangle\langle\psi_{i}|$ attains the minimal decomposition of the multipartite concurrence, one has,

$$C_{5}^{2}(\rho) = (\sum_{i}p_{i}C_{5}(|\psi_{i}\rangle\langle\psi_{i}|))^{2}~~~~~~~~~~~~~~~~~~~~~~~~~~~~~~~~~~~~~~~~~~~~~~~~~~~~~~~~~~~~~~~~~~~~~~~~~~~~~~~~~~~$$
$$\geq (\sum_{i}p_{i}\sqrt{\frac{1}{25}(C_3^{2}((|\psi_{i}\rangle)_{1|2|345})+C_3^{2}((|\psi_{i}\rangle)_{1|3|245})+\cdots+C_3^{2}((|\psi_{i}\rangle)_{145|2|3}))})^{2}~~~~~$$
$$~~~~~~~~~~~\geq (\sum_{i}p_{i}\frac{1}{5}C_3((|\psi_{i}\rangle)_{1|2|345}))^{2} + (\sum_{i}p_{i}\frac{1}{5}C_3((|\psi_{i}\rangle)_{1|3|245}))^{2} + \cdots + (\sum_{i}p_{i}\frac{1}{5}C_3((|\psi_{i}\rangle)_{145|2|3}))^{2}$$
$$~~~~~~~~~~~\geq \frac{1}{25}(C_3^{2}(\rho_{1|2|345})+C_3^{2}(\rho_{1|3|245})+\cdots+C_3^{2}(\rho_{145|2|3})),~~~~~~~~~~~~~~~~~~~~~~~~~~~~~~~~~~~~~~~~~~~~~~~~~~~~~~~~~~~~~~~~~~~~~~~~~~~~$$

\noindent Therefore, we have (\ref{13}).

If we take $N=5, M=4,$ there are ten different
 partitions of five system: $1|2|3|45, 1|2|4|35, 1|2|5|34,$

\noindent$1|23|4|5, 1|24|3|5, 1|25|3|4, 12|3|4|5, 13|2|4|5, 14|2|3|5, 15|2|3|4,$ similar to Theorem2, we can get

\noindent{\bf Theorem 3.} For a multipartite quantum state $\rho \in  H_1\otimes H_2\otimes H_3 \otimes H_4\otimes H_5,$  then the following inequality holds,
\begin{eqnarray}\label{18}
C_5^{2}(\rho) \geq  \widetilde{C_4}^{2} (\rho),
\end{eqnarray}
where $\widetilde{C_4}^{2} (\rho) = \frac{1}{10}(C_4^{2}(\rho_{1|2|3|45})+C_4^{2}(\rho_{1|2|4|35})+\cdots + C_4^{2}(\rho_{15|2|3|4})).$

\noindent \emph{Proof:}  For a pure multipartite state $|\psi\rangle \in H_1\otimes H_2\otimes H_3 \otimes H_4\otimes H_5,$ let $\rho = |\psi\rangle\langle\psi|,$
From (\ref{1}), we have

\begin{eqnarray}\label{19}
C_5^{2}(\rho)=\frac{1}{4} (\sum_{i=1}^{5}(1-tr\rho^{2}_{i}) + \sum_{i=2}^{5}(1-tr\rho^{2}_{1i})+ \sum_{i=3}^{5}(1-tr\rho^{2}_{2i})+ \sum_{i=4}^{5}(1-tr\rho^{2}_{3i})+ (1-tr\rho^{2}_{45})),
\end{eqnarray}

and

\begin{eqnarray}\label{20}
\widetilde{C_4}^{2} (\rho) = \frac{1}{10}(C_4^{2}(\rho_{1|2|3|45})+C_4^{2}(\rho_{1|2|4|35})+\cdots + C_4^{2}(\rho_{15|2|3|4}))~~~~~~~~~~~~~~~~~~~~~~~~~~~~~~~~~~~
\nonumber \\= \frac{1}{20}\{[(1-tr\rho_{12}^2)+(1-tr\rho_{3}^2)+(1-tr\rho_{4}^2)+(1-tr\rho_{5}^2)+(1-tr\rho_{34}^2)~~~~~~~~~~~~~~~
\nonumber \\+(1-tr\rho_{35}^2)+(1-tr\rho_{45}^2)]+[(1-tr\rho_{13}^2)+(1-tr\rho_{2}^2)+(1-tr\rho_{4}^2)~~~~~~~~~~~~~~~~~~
\nonumber \\+(1-tr\rho_{5}^2)+(1-tr\rho_{45}^2)+(1-tr\rho_{25}^2)+(1-tr\rho_{24}^2)]+\cdots+[(1-tr\rho_{45}^2)~~~~~~~~~
\nonumber \\+ (1-tr\rho_{1}^2)+(1-tr\rho_{2}^2)+(1-tr\rho_{3}^2)+(1-tr\rho_{23}^2)+(1-tr\rho_{13}^2)+(1-tr\rho_{12}^2)]\}
\end{eqnarray}

In order to prove (\ref{18}) for pure state, we compare (\ref{19}) with (\ref{20}), and we find that we only to prove
\begin{eqnarray}\label{21}
\sum_{i=1}^{5}(1-tr\rho^{2}_{i})\leq \sum_{i=2}^{5}(1-tr\rho^{2}_{1i})+ \sum_{i=3}^{5}(1-tr\rho^{2}_{2i})+ \sum_{i=4}^{5}(1-tr\rho^{2}_{3i})+ (1-tr\rho^{2}_{45})
\end{eqnarray}
and (\ref{21}) is obvious right for the pure state by (\ref{17}).

Assuming that a mixed state $\rho = \sum_{i}p_{i}|\psi_{i}\rangle\langle\psi_{i}|$ attains the minimal decomposition of the multipartite concurrence, one has,

$$C_{5}^{2}(\rho) = (\sum_{i}p_{i}C_{5}(|\psi_{i}\rangle\langle\psi_{i}|))^{2}~~~~~~~~~~~~~~~~~~~~~~~~~~~~~~~~~~~~~~~~~~~~~~~~~~~~~~~~~~~~~~~~~~~~~~~~~~~~~~~~~~~~~~~~~~~~~~~~~~$$
$$\geq (\sum_{i}p_{i}\sqrt{\frac{1}{10}(C_4^{2}((|\psi_{i}\rangle)_{1|2|3|45})+C_4^{2}((|\psi_{i}\rangle)_{1|2|4|35})+\cdots+C_4^{2}((|\psi_{i}\rangle)_{15|2|3|4}))})^{2}~~~~~~~~~~~~~~~~~~~~~~~$$
$$\geq (\sum_{i}p_{i}\frac{1}{\sqrt{10}}C_4((|\psi_{i}\rangle)_{1|2|3|45}))^{2} + (\sum_{i}p_{i}\frac{1}{\sqrt{10}}C_4((|\psi_{i}\rangle)_{1|2|4|35}))^{2} + \cdots + (\sum_{i}p_{i}\frac{1}{\sqrt{10}}C_4((|\psi_{i}\rangle)_{15|2|3|4}))^{2}$$
$$\geq \frac{1}{10}(C_4^{2}(\rho_{1|2|3|45})+C_4^{2}(\rho_{1|2|4|35})+\cdots+C_4^{2}(\rho_{15|2|3|4})),~~~~~~~~~~~~~~~~~~~~~~~~~~~~~~~~~~~~~~~~~~~~~~~~~~~~~~~~~~~~~~~~~~~~~~~~~~~~~~~~~~~~~~~~~~~~~~~~~~~~~~~~~~~~~~~~~~~~~~~~~~~~~~~~~~~~~~~~~~~~~~~~~~~~$$

\noindent Therefore, we have (\ref{18}).

Now we hope to generalize our results to $N$-partite systems ($N>4$). Firstly, we consider six-partite state $\rho \in  H_1\otimes H_2\otimes H_3 \otimes H_4\otimes H_5\otimes H_6,$ if we hope to get $C_6^{2}(\rho) \geq  \widetilde{C_5}^{2} (\rho),$ we should get
\begin{eqnarray}\label{22}
5\sum_{i=1}^{6}(1-tr\rho^{2}_{i})\leq \sum_{i=2}^{6}(1-tr\rho^{2}_{1i})+ \sum_{i=3}^{6}(1-tr\rho^{2}_{2i})+ \sum_{i=4}^{6}(1-tr\rho^{2}_{3i})+ \sum_{i=5}^{6}(1-tr\rho^{2}_{4i})
\nonumber \\+ (1-tr\rho^{2}_{56})+3[\sum_{i=3}^{6}(1-tr\rho^{2}_{12i})+\sum_{i=4}^{6}(1-tr\rho^{2}_{13i})+\sum_{i=5}^{6}(1-tr\rho^{2}_{14i})+(1-tr\rho^{2}_{156})],
\end{eqnarray}
but we are not sure that (\ref{22}) is always true. So we only have $C_{N}^{2}(\rho) \geq \widetilde{C_{M}}^{2}(\rho)$ for integers $N\leq 5, 3\leq M \leq4$, where $\widetilde{C_{M}}^{2}(\rho)$  takes average over all possible square $M$-partite concurrences. Generally, we obtain

\noindent{\bf Theorem 4.}
$$
C_{5}^{2}(\rho) \geq
s_1\{\widetilde{C_{4}}^{2}(\rho)\}+s_2\{\widetilde{C_{3}}^{2}(\rho)\},
$$
where $\sum_{i=1}^{2}s_i = 1$, $s_1\geq 0, s_2\geq 0.$

\section{Conclusions and Remarks}\label{sec3}
In summary, we have presented an approach to derive lower bounds of concurrence for arbitrary dimensional $N$-partite ($N \leq 5$) systems based on sub $M$-partite ($M=3,...,N-1$) concurrences.
Lower bounds of concurrence for four-partite(or five-partite) mixed states have been studied in detail in terms of the tripartite concurrences.
By detailed examples we have shown that this bound is better than other existing lower bounds of concurrence.
we find that our lower bound is relatively tight though these examples. Example 1 in our paper can show that our lower bound are larger than the previous lower bounds and Example 2, 3 can show our lower bound can detect more entanglement than the previous lower bounds. At last, we also present a lower bound of five-partite mixed states based on the mixing of $\widetilde{C_{4}}^{2}(\rho)$ and $\widetilde{C_{3}}^{2}(\rho)$.

Above all, in \cite{Fan,Zhu0,Ch} lower bounds of concurrence for high dimensional systems have been presented based on
the concurrences of sub-dimensional states, by decomposing
the joint Hilbert space into lower dimensional subspaces.
For high dimensional $N$-partite systems ($N\leq 5$), it would be
useful to use the concurrences of both sub-dimensional states and sub-partite states.
An optimal lower bound could be obtained by repeatedly using
the concurrences of sub-dimensional and sub-partite states in an suitable order. We are also looking forward to get lower bounds for arbitrary multipartite systems.

\vspace{2.5ex}
\noindent{\bf Acknowledgments}\, \,
This work is supported by the NSFC under numbers 11571119 and 11405060.


\begin{thebibliography}{}

\bibitem{Nie} Nielsen, M.A., Chuang, I.L.: Quantum Computation and Quantum Information(Cambridge University Press, Cambridge, 2000).

\bibitem{Ein} Einstein, A., Podolsky, B., Rosen, N.: Can Quantum-Mechanical Description of Physical Reality Be Considered Complete?. Phys. Rev. 47, 777(1935).

\bibitem{Ami} Osterloh, A., Amico, L., Falci, G., Fazio, R.: Scaling of entanglement close to a quantum phase transition. Nature(London)416, 608(2002).

\bibitem{Wer} Werner, R.F.: Quantum states with Einstein-Podolsky-Rosen correlations admitting a hidden-variable model. Phys. Rev. A 40, 4277(1989).

\bibitem{Hor} Horodecki, R., Horodecki, P., Horodecki, M., Horodecki, K.: Quantum entanglement. Rev. Mod. Phys. 81, 865(2009).

\bibitem{Flo} Mintert, F., Ku$\acute{s}$, M., Buchleitner, A.: Concurrence of mixed bipartite quantum states in arbitrary dimensions. Phys. Rev. Lett. 92, 167902(2004);\\
 Mintert, F.: Measures and dynamics of entangled states, Ph.D. thesis, Munich University, Munich, 2004.

\bibitem{Chen} Chen, K., Albeverio, S., Fei, S.M.: Concurrence of arbitrary dimensional bipartite quantum states. Phys. Rev. Lett. 95, 040504(2005).

\bibitem{Bre} Breuer, H.P.: Separability criteria and bounds for entanglement measures. J. Phys. A: Math. Gen. 39, 11847 (2006).

\bibitem{Bre1} Breuer, H.P.: Optimal entanglement criterion for mixed quantum states. Phys. Rev. Lett. 97, 080501(2006).

\bibitem{Vic} de Vicente, J.I.: Lower bounds on concurrence and separability conditions. Phys. Rev. A 75, 052320(2007).

\bibitem{di} DiVincenzo, D.P.: Quantum Computation. Science 270, 5234(1995).

\bibitem{teleportation}
 Bennett, C.H., Brassard, G., Cr\'{e}peau, C., Jozsa, Peres  R.A., Wootters, W.K.: Teleporting an unknown quantum state via dual classical and Einstein-Podolsky-Rosen channels. Phys. Rev. Lett. 70, 1895 (1993);\\
Albeverio, S., Fei, S.M., Yang, W.L.: Optimal teleportation based on bell measurements. Phys. Rev. A 66, 012301(2002).

\bibitem{dense} Bennett C.H., Wiesner, S.J.: Communication via one- and two-particle operators on Einstein-Podolsky-Rosen states. Phys. Rev. Lett. 69, 2881(1992).

\bibitem{schemes} Ekert, A.: Quantum cryptography based on Bell¡¯s theorem. Phys. Rev. Lett. 67, 661(1991);\\
Fuchs, C.A., Gisin, N., Griffiths, R.B., Niu, C.S., Peres, A.: Optimal eavesdropping in quantum cryptography. I. Information bound and optimal strategy. Phys. Rev. A 56, 1163(1997).

\bibitem{swapping} \.{Z}ukowski, M., Zeilinger, A., Horne M.A., Ekert, A.K.: ¡®¡®Event-ready-detectors¡¯¡¯ Bell experiment via entanglement swapping. Phys. Rev. Lett. 71, 4287(1993);\\
 Bose, S., Vedral V., Knight, P.L.: Multiparticle generalization of entanglement swapping. Phys. Rev. A 57, 822(1998); 60, 194(1999);\\
Shi, B.S., Jiang, Y.K., Guo, G.C.: Optimal entanglement purification via entanglement swapping. Phys. Rev. A 62, 054301(2000).

\bibitem{RSP1} Bennett, C.H., DiVincenzo, D P., Shor, P.W., Smolin, J.A.,
Terhal B.M., Wootters, W.K.: Remote State Preparation. Phys. Rev. Lett. 87, 077902(2001);\\
Leung D.W., Shor, P.W.: Oblivious Remote State Preparation. Phys. Rev. Lett. 90,
127905(2003).

\bibitem{Ben2} Bennett, C.H., DiVincenzo, D.P., Smolin, J.A., Wootters, W.K.: Mixed-state entanglement and quantum error correction. Phys. Rev. A 54,3824(1996);\\
Plenio M.B., Virmani, S.: An introduction to entanglement measures. Quant. Inf. Comput. 7, 1(2007).

\bibitem{Uhl} Uhlmann, A.: Fidelity and concurrence of conjugated states. Phys. Rev. A 62, 032307(2000);\\
Rungta, P., Bu$\check{z}$ek,  V., Caves, C.M., Hillery, M., Milburn, G.J.: Universal state inversion and concurrence in arbitrary dimensions. Phys. Rev. A 64, 042315(2001);\\
 Albeverio S., Fei, S.M.: A note on invariants and entanglements. J. Opt. B: Quantum Semiclassical Opt. 3, 223(2001).

\bibitem{Woo} Wootters, W.K.: Entanglement of formation of an arbitrary state of two qubits. Phys. Rev. Lett. 80, 2245(1998).

\bibitem{Ter} Terhal, B.M., Vollbrecht, K.G.H.: Entanglement of formation for isotropic states. Phys. Rev. Lett. 85, 2625(2000)\\
Fei, S.M., Jost, J., Li-Jost, X.Q., Wang, G.F.: Entanglement of formation for a class of quantum states. Phys. Lett. A 310, 333(2003);\\
Rungta P., Caves, C.M.: Concurrence-based entanglement measures for isotropic states. Phys. Rev. A 67, 012307(2003);\\
Fei, S.M., Li-Jost, X.Q.: A class of special matrices and quantum entanglement. Rep. Math. Phys. 53, 195(2004);\\
Fei, S.M., Wang, Z.X., Zhao, H.:  A note on entanglement of formation and generalized concurrence. Phys. Lett. A 329, 414(2004).

\bibitem{Zha} Zhang, C.J., Zhang, Y.S., Zhang, S., Guo, G.C.: Optimal entanglement witnesses based on local orthogonal observables. Phys. Rev. A 76, 012334(2007).

\bibitem{Ger} Gerjuoy, E.: Lower bound on entanglement of formation for the qubit-qudit system. Phys. Rev. A 67,052308(2003).

\bibitem{Fan} Ou, Y.C., Fan, H., Fei, S.M.: Proper monogamy inequality for arbitrary pure quantum states. Phys. Rev. A 78, 012311(2008).

\bibitem{Zhu} Zhao, M.J., Zhu, X.N., Fei, S.M., Li-Jost, X.Q.: Lower bound on concurrence and distillation for arbitrary-dimensional bipartite quantum states. Phys. Rev. A 84, 062322(2011).

\bibitem{Zhu0} Zhu, X.N.,Zhao, M.J., Fei, S.M.: Lower bound of multipartite concurrence based on subquantum state decomposition. Phys. Rev. A 86, 022307(2012).
    
\bibitem{Ch} Chen, W., Fei, S.M., Zheng, Z.J.:  Lower bound on concurrence for arbitrary-dimensional tripartite quantum states. Quantum Inform. Processing 15, 3761-3771(2016).

\bibitem{Wang} Li, M., Fei, S.M., Wang, Z.X.: Bounds for multipartite concurrence. Rep. Math. Phys. 65, 289-296(2010).

\bibitem{Zhu1} Zhu, X.N., Li, M., Fei, S.M.: Lower bounds of concurrence for multipartite states. Aip Conf. Proc. Advances in Quan. Theory, 1424(2012).

\bibitem{Aol} Aolita L., Mintert, F.: Measuring Multipartite Concurrence with a Single Factorizable Observable. Phys. Rev. Lett. 97, 050501(2006);\\
Carvalho, A.R.R., Mintert, F., Buchleitner, A.: Decoherence and Multipartite Entanglement. Phys. Rev. Lett. 93, 230501(2004).

\bibitem{Zhu2} Zhu, X.N., Fei, S.M.: Lower bound of concurrence for qubit systems. Quantum Inform. Processing 13, 815-823(2014).

\bibitem{Qi} Qi, X.F., Gao, T., Yan, F.l.: Lower bounds of concurrence for $N$-qubit systems and the detection of $k$-nonseparability of multipartite quantum systems. arXiv:1605.05000(2016).

\bibitem{Kie} Kiesel, N., Schmid, C., T$\acute{o}$th, G., Solano, E., Weinfurter, H.: Experimental Observation of Four-Photon Entangled Dicke State with High Fidelity. Phys. Rev. Lett. 98, 063604(2007).

\end{thebibliography}
\end{document}